# Entanglement generation by Fock-state filtration

K. J. Resch[1], J. L. O'Brien[1†], T. J. Weinhold[1], K. Sanaka[2], and A. G. White[1]

[1]*Department of Physics, Centre for Quantum Computer Technology, University of Queensland, Brisbane, Australia*
[2]*E. L. Ginzton Laboratory, Stanford University, Stanford, USA*

We demonstrate a Fock-state filter which is capable of preferentially blocking single photons over photon pairs. The large conditional nonlinearities are based on higher-order quantum interference, using linear optics, an ancilla photon, and measurement. We demonstrate that the filter acts coherently by using it to convert unentangled photon pairs to a path-entangled state. We quantify the degree of entanglement by transforming the path information to polarisation information, applying quantum state tomography we measure a tangle of $T=(20 \pm 9)\%$.

In practice it is extremely difficult to make one photon coherently influence the state of another. The optical nonlinearities required are orders of magnitude beyond those commonly achieved in current photonic technology. Fortunately, strong effective nonlinearities can be induced in linear optical systems by combining quantum interference and projective measurement [1], opening the possibility of scalable linear-optical quantum computation. Such *measurement-induced nonlinearities* have had high impact in quantum information and optics, notably in optical quantum logic gate experiments [2–5] and in exotic state production [6–9].

Most of these schemes achieve an effective nonlinearity via the lowest-order nonclassical interference, with one photon per mode input to a beamsplitter. Higher-order nonclassical interference, where more than one photon is allowed per mode, enables additional control over the quantum state [1]. A single ancilla photon has been used to conditionally control the phase of a two photon path-entangled state [2], and to conditionally absorb either one- or two-photon input states [10]. Applied to a superposition state, higher-order interference is predicted to act as a Fock-state filter [11–14], conditionally absorbing only terms with a specified number of photons. In this paper, we prove that this conditional absorption is *coherent* by applying it to a quantum superposition, and experimentally generating a path-entangled state. We quantify the degree of entanglement by transforming the path information to polarisation information, and applying quantum state tomography [15].

The Fock state filter is based on a non-classical interference at a single, polarisation-independent, beamsplitter of reflectivity, $R$. Consider the beamsplitter in the Fock-state filter of Fig. 1 with $n+1$ photons incident: $n$ in mode $a$, and 1 in mode $b$, the latter labelled the ancilla photon. There are $n+1$ possible ways for there to be one and only one photon in mode $d$: all the input photons can be reflected, with probability amplitude $\sqrt{R}^{n+1}$, or there are $n$ ways for a photon from each input to be transmitted and the rest reflected, $n(1-R)\sqrt{R}^{n-1}$. Assuming indistinguishable photons, the probability amplitude for detecting one and only one photon in mode $d$ is [10, 11, 16]:

$$A(n) = R^{(n-1)/2}[R - n(1 - R)], \quad (1)$$

Note that the probability, $P(n)=|A(n)|^2$, can be zero for any single choice of $n$, when $R=n/(n + 1)$; for all other $n$, $P>0$ [10]. Hong-Ou-Mandel interference is the lowest-order case, where $P=0$ when $n=1$ and $R=1/2$ [17]: the detector in mode $d$ is never hit by a single photon. Consider the action of the filter when a coherent superposition of number states is input into mode $a$. If a single photon is detected in mode $d$ then the output state in mode $c$ cannot contain a single photon, $|1\rangle$.

In principle, a Fock-state filter can be tested by creating a number-state superposition in one spatial mode, applying the filter to it, and tomographically measuring the resulting state. In practice, each step of this naïve approach is impractical: creating non-classical number-state superpositions is onerous [6, 7, 18–21]; it is difficult to maintain such states as they are easily destroyed by the loss of even one photon; the Fock-state filter requires an ancilla photon on demand and a perfect-efficiency number-resolving detector; and the tomographic reconstruction needs high-efficiency homodyne measurement.

We designed our experiment to alleviate each of these difficulties. We use *double-pair* emission events to generate a pair of polarised two-photon states in separate spatial modes. It is often pointed out that parametric down-conversion is a problematic source of photon pairs, since it emits them probabilistically, and can emit more than one pair at a time. In some cases, this double-pair emission is either beneficial (e.g., entanglement purification [22–24]), or indeed, essential (e.g., multi-photon entangled states [25, 26]). Double-pair emission provides us with input two-photon states in mode $a$ and single, ancillary, photons in mode $b$: we create the superposition in mode $a$ by rotating its polarisation,

$$|2_H, 0_V\rangle_a \rightarrow \cos^2\theta |2_H, 0_V\rangle_a + \sin^2\theta |0_H, 2_V\rangle_a$$
$$+ \sqrt{2}\cos\theta\sin\theta |1_H, 1_V\rangle_a, \quad (2)$$

where $\theta$ is the polarisation angle relative to horizontal. We create a horizontally-polarised ancilla photon in mode

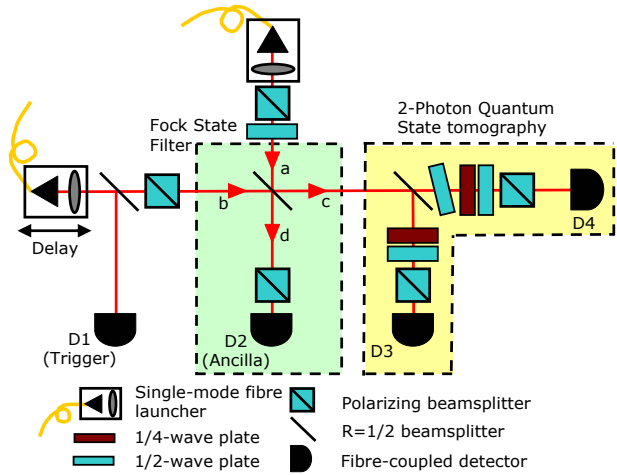

FIG. 1: The Fock state filter: a device that completely blocks the passage of single photons, but allows the coherent passage of photon pairs, and so does not destroy superposition states. As described in the text, a probabilistic Fock-state filter can be created by combining a 50% beamsplitter, ancilla photon, quantum interference, and measurement.

$b$ by passing the two-photon state through a 50% beam-splitter and triggering on detection events from the output mode of the beamsplitter, see Fig. 1. The trigger photon is measured in coincidence with the three photons output from the beamsplitter: if a photon is lost anywhere in the experiment then a four-fold coincidence is not registered and therefore the experimental signal is not susceptible to photon loss.

The Fock-state filter acts nonlinearly only on light with the same polarisation as the ancilla, horizontal in this case. The amplitude given in equation 1 determines the transformation on horizontally-polarised components of the state,

$$|n_\mathrm{H}\rangle|1_\mathrm{H}\rangle \rightarrow A(n_\mathrm{H})|n_\mathrm{H}\rangle|1_\mathrm{H}\rangle + ... \quad (3)$$

In contrast, the vertically-polarised components are transformed as,

$$|n_\mathrm{V}\rangle|1_\mathrm{H}\rangle \rightarrow R^{(n_\mathrm{V}+1)/2}|n_\mathrm{V}\rangle|1_\mathrm{H}\rangle + ... \quad (4)$$

Measurement of a single horizontally-polarised photon in mode $d$ selects only the first term of Eqns 3 and 4 (the latter amplitude represents the only way that a horizontally-polarised photon can be detected in mode $d$). Noting that the conditional transformation is not unitary, and applying this to the terms in equation 2 we find,

$$|2_H, 0_V\rangle_a \rightarrow -\frac{1}{2\sqrt{2}}|2_H, 0_V\rangle_c \quad (5)$$

$$|1_H, 1_V\rangle_a \rightarrow 0 \quad (6)$$

$$|0_H, 2_V\rangle_a \rightarrow \frac{1}{2\sqrt{2}}|2_H, 0_V\rangle_c, \quad (7)$$

and thus the state of mode $c$ conditioned on a horizontal photon detected in mode $d$ is,

$$\frac{-\cos^2\theta|2_H, 0_V\rangle_c + \sin^2\theta|0_H, 2_V\rangle_c}{(\cos^4\theta + \sin^4\theta)^{1/2}}. \quad (8)$$

The final state can be tuned between separable and entangled number-path states simply by adjusting the input polarisation, $\theta$. In the case, $\theta=\pi/4$, this is the lowest-order NOON state [27], $(|2_H, 0_V\rangle-|0_H, 2_V\rangle)/\sqrt{2}$.

Note that the vertical polarisation provides an intrinsically stable phase reference for the nonlinear sign change of the horizontal components, removing the need for an actively-stabilised homodyne measurement. The final state is transformed from one to two spatial modes by a 50% beamsplitter: mapping the path-entanglement into polarisation-entanglement lets us characterise the state with quantum state tomography of the polarisation, with all of its attendant advantages [15].

Our downconversion source was a BBO ($\beta$-barium borate) nonlinear crystal cut for noncollinear type-I frequency conversion (410 nm→820 nm), pumped by a frequency-doubled Titanium Sapphire laser. The downconverted light was coupled into two single-mode optical fibres, which when connected directly to FC-connectorised single-photon counting modules yielded coincidence rates of 30 kHz and singles rates of 220 kHz. Before coupling back into free-space, the polarisation of the light was manipulated in-fibre using "bat-ears" to maximise transmission through horizontal polarisers. Light in mode $b$ was split by a 50% beamsplitter, where one output mode was coupled directly into a single-mode fibre coupled detector, D1, which acts as a trigger. The remaining light passed through a horizontal polarizer and is combined on a second 50% beamsplitter with light from mode $a$, which is first passed through a horizontal polarizer and half-wave plate to rotate the polarisation, as described in Eq. 2. Mode $d$ is directly detected at D2; mode $c$ is split into two modes by a 50% beamsplitter, each mode is polarisation analysed using a quarter- and half- wave plate and polarizer. We use D3 and D4 to perform a tomographically-complete set of two-qubit measurements, $\{H, V, D, R\} \otimes \{H, V, D, R\}$, in coincidence with the trigger and ancilla detectors, D1 & D2. The resulting density matrices are reconstructed using the maximum-likelihood technique [15]. All of the optical paths between fibre couplers to detectors were made approximately equal ($\sim$50cm) to make possible high-efficiency single-mode to single-mode fibre coupling. The tilted half- wave plate in the D4 arm, set with its optic axis horizontal, was used to compensate birefringence in the beamsplitters.

Nonclassical interference is the heart of the Fock-state filter. We characterised this by setting the polarisation of mode $a$ to horizontal, matching that of mode $b$, and setting analysers at D3, D4 to horizontal. Fig. 2 shows experimentally measured two-fold coincidence counts, in this case between detectors D2 & D4 (open circles), and the four-fold coincidence counts, between D1, D2, D3, & D4 (solid circles), as a function of the longitudinal position of the input fibre coupler for mode $b$.

As D2 and D4 detect the two outputs of the beamsplit-ter, the two-folds show the standard Hong-Ou-Mandel interference dip [17], with a raw visibility of $V_1$=(95.20 ± 0.02)%. This does not yet suffice to allow us to estimate the performance of the Fock-state filter: the two-fold rate has significant contributions from the two-photon terms in modes $a$ and $b$. We can estimate these by blocking mode $a$ and $b$ in turn and measuring the the two-fold coincidences between detectors D2 & D4, 5.8 ± 0.16 Hz and 30.9 ± 0.5 Hz, respectively. Summing these gives an estimate of the number of two-fold coincidences due to the the two-photon terms in modes $a$ and $b$, (36.7 ± 0.5) Hz. These coincidences act as a background in the nonclassical visibility, subtracting them gives a corrected visibility of $V_1'$=(99.6 ± 0.1)%.

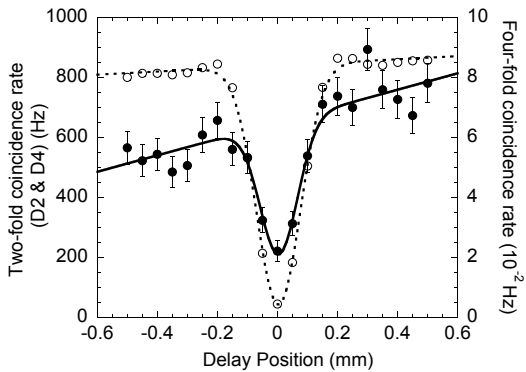

FIG. 2: High-visibility quantum interference in two- and four-fold coincidence counts as a function of the longitudinal position of the input fibre coupler for mode $b$. At zero delay, we see marked preferential absorption of single-photon over two-photon states in mode $c$, as indicated by the larger dip in two- over four- fold counts. The two- and four- fold raw visibilities are (95.20±0.02)% and (68±5)%, respectively; correcting for background as described in the text, the two-fold visibility becomes, (99.6±0.1)% (error bars are smaller than the points in the two-fold case and are not shown). The visibilities are in excellent agreement with the theoretically expected two- and four- visibilities of 100% and 66.7% [10, 17, 28]. The input coupler was scanned 1 mm in 630 s: to mitigate drift effects the scan was repeated 63 times, leading to an integration time of 31.5 minutes per point. The slopes in the data are due to longitudinal-position-dependent coupling to the detectors; the trigger-detector was particularly sensitive in this respect, leading to a large slope in the four-folds; the two-folds show a much smaller slope as the trigger detector plays no role in that data. The visibilities were obtained from curve fits to products of a Gaussian and a linear function.

The four-fold coincidence counts in Fig. 2 display a higher-order nonclassical interference effect, as expected from Eq. 1. Our visibility is $V_2$=(68±5)%, which agrees with the expected value of 66.7% [10]. Note that the interference visibility is much larger for the $n$=1 input state, as measured by the two-fold coincidences, than the $n$=2 input state, as measured by the four-fold coincidences. At the centre of the interference dip, single photons are removed from an input state with much higher probability than pairs of photons: this is the action of the Fock-state filter.

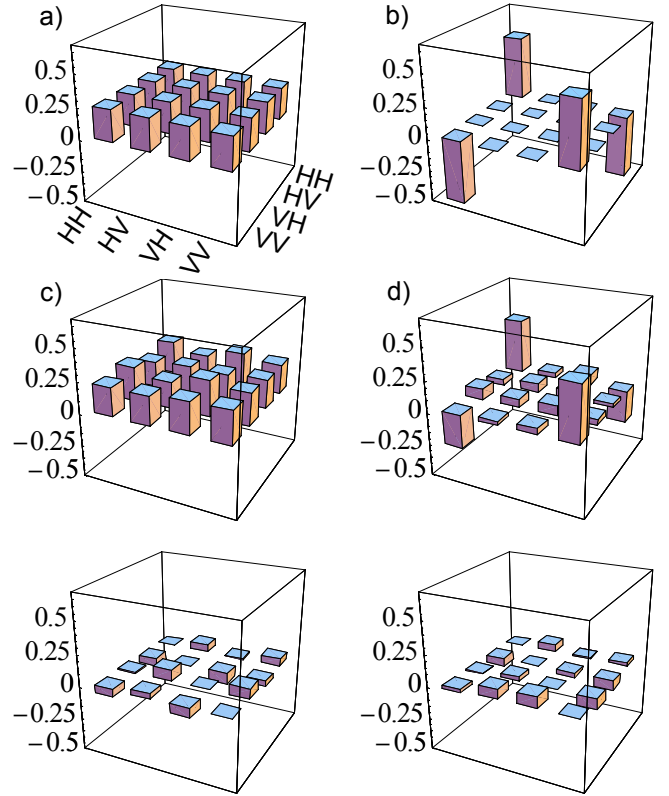

FIG. 3: Density matrices for the Fock-state filter. Ideal output states from the filter when the filtering is a) turned off, $|DD\rangle$, & b) turned on, $(|HH\rangle-|VV\rangle)/\sqrt{2}$, as described in text. The corresponding experimental tomographic reconstructions, based on raw counts, are shown respectively in c) & d), the upper panels are the real components, the lower panels, imaginary. The fidelity between the ideal and measured states is 93 ± 4% and 69 ± 9%, respectively. The state measured in d) is entangled, with tangle $T = 20 \pm 9$%.

The visibilities, $V_1'$ & $V_2$, allow us to set an upper bound to the performance of the Fock-state filter. Ideally, the probability of transmission when the ancilla and $n$-photon inputs are distinguishable is $Q(n)$=$R^{n+1}+nR^{n-1}(1-R)^2$ [10]. The nonlinear absorption probability, $P(n)$, is modified by the visibilities as $P'(n)$=$(1-V_n)Q(n)$. We use this to estimate the efficiency of the nonlinear absorber at blocking the passage of single photons, $P'(2)/P'(1)$=60 ± 20, that is the Fock-state filter will, at best, preferentially pass two-photon terms at 60 times the rate it passes single-photon terms.

To show the coherent action of the Fock-state filter, we set the input waveplate in mode $a$ to rotate the linear polarisation from horizontal to diagonal, $|D\rangle$=$(|H\rangle+|V\rangle)/\sqrt{2}$, creating the superposition of Eq. 2. We first measure the input state without the action of the Fock state filter by blocking the ancilla photon in mode $b$, and performing tomography on mode $c$ using detectors D3 & D4. Counting for 30 s per measurement setting, we measured raw two-fold coincidence counts of {86, 68, 156, 61,

89, 77, 195, 61, 200, 170, 328, 131, 98, 102, 175, 71}. The reconstructed density matrix, shown in Fig. 3c), gives us the initial state of the light and includes the effect of any birefringence in our experiment. The density matrix consists of near equal probabilities, and strong positive coherences between them—characteristic of the expected ideal state $|\psi\rangle=|DD\rangle$. The fidelity between the ideal state and the measured state, $\rho$, is $\mathcal{F}=\langle\psi|\rho|\psi\rangle=(93\pm4)\%$; the linear entropy is $S_L=(11\pm8)\%$ [15], indicating the state is near-pure; and the tangle is zero within error, $T=(0.5\pm0.8)\%$, indicating that as expected the input state is unentangled.

The Fock-state filter is run by unblocking mode $b$ and setting its coupler to the zero-delay position shown in Fig. 2. As before, we performed tomography on the photon pairs at D3 & D4, but now in coincidence with the trigger and ancilla photon detectors, D1 & D2, counting for 8.25 hours per measurement setting, obtaining the raw counts {62, 10, 45, 25, 10, 59, 49, 49, 53, 40, 36, 45, 37, 50, 46, 72}. (Note that measuring four-folds at the bottom of a high-visibility non-classical interference dip leads to a very low count rate!). The reconstructed density matrix is shown in Fig. 3d). Consistent with the prediction of Eq. 8 setting $\theta=\pi/4$, there are two striking differences between this and Fig. 3c): 1) the dramatic reduction of the HV and VH populations and their associated coherences; and 2) the sign change of the coherences between the HH and VV populations. The fidelity, between the ideal state, $|\psi\rangle=(|HH\rangle-|VV\rangle)/\sqrt{2}$, and the measured state, $\rho$, is $\mathcal{F}=(69\pm9)\%$. The linear entropy is $S_L=(57\pm6)\%$, the increase in entropy indicates that the Fock-state filter introduces some mixture but retains much of the coherence of the input state. This is reflected by the output state, which is clearly entangled, $T=(20\pm9)\%$, indicating a coherent superposition of output states.

The tomography is based on the four-fold signal, which is particularly susceptible to background counts, due to the combination of low rate and long counting times. We use raw, rather than corrected four-fold counts, as unambiguous measurement of the background is non-trivial due to the manifold combinations of accidental detection events. Thus $(P_{HH}+P_{VV})/(P_{HV}+P_{VH})$ is very much a lower bound to the preferential absorption of our Fock-state filter: from our counts we measure $6.0\pm1.5$.

Although initially invented in the context of optical quantum computation, measurement-induced nonlinearities have enormous potential throughout quantum optics. Here we have constructed a coherent nonlinear absorber—a Fock-state filter—combining measurement with higher-order quantum interference. The filter preferentially absorbed up to 60 times more single photons than photon pairs, and was used to produce an entangled state from an separable input state: using quantum tomography, the output was measured to have a tangle of $T=(20\pm9)\%$. By encoding quantum information in both number and polarisation, and moving between number and polarisation entanglement, we were able to succinctly demonstrate all the salient features of a Fock-state filter in a single experiment. This is a powerful technique suitable for applications requiring *quantum* nonlinear optics.

We thank Anton Zeilinger for valuable discussions. This work was supported in part by the DTO-funded U.S. Army Research Office Contract No. W911NF-05-0397, a University of Queensland Early Career Researcher Grant, and the Australian Research Council Discovery program.

[†]Department of Electrical & Electronic Engineering, University of Bristol, BS8 1UB, UK.